\documentclass[aps,prl,twocolumn,groupedaddress,eqsecnum,showpacs]{revtex4-1}
\usepackage{amsfonts}
\usepackage{dcolumn}
\usepackage{graphicx,amssymb,amsmath}
\usepackage{bm}
\usepackage{natbib}
\usepackage{epstopdf}
\begin{document}

\title{Quantum phase coherence in non-Markovian and reaction-diffusive transport}

\author{Shimul Akhanjee}
\affiliation{Condensed Matter Theory Laboratory, RIKEN, Wako, Saitama, 351-0198, Japan}
\email[]{shimul@riken.jp}
\date{\today}

\begin{abstract}
We study quantum phase coherence and weak localization (WL) in disordered metals with restricted back-scattering and phenomenologically formulate a large class of unconventional transport mechanisms as modified diffusion processes not captured by the Boltzmann picture. Inspired by conductivity measurements in ferromagnetic films and semiconductors where anomalous power law corrections have been observed, we constrain memory dependent, self avoidance effects onto the quantum enhanced back-scattered trajectories, drastically altering the effect of weak localization in two dimensions (2D). Scale dependent corrections to the conductivity fail to localize the electrons in $d \ge 2$ for sufficiently weak disorder. Additionally, we analyze quantum transport in reaction-diffusion systems governed by the Fisher's equation and observe asymptotically similar delocalization in 2D. Such unconventional transport might be relevant to certain non-Fermi liquid or strongly correlated phases in 2D within the negative compressibility regime.  
  
\end{abstract}



\maketitle

\emph{Introduction}--
Lacking an adequate mathematical theory of the transition point and no rigorous proof of the existence of extended states in three dimensions (3D), Anderson localization of non-interacting electrons is an ongoing endeavor of condensed matter physics\cite{anderson}. Modern approaches exploit symmetry classes of a given disordered Hamiltonian, constructing perturbative conductance $g$ scaling forms of the diffusion (Goldstone) modes. 10 such universality classes have been identified with certain cases demonstrating an absence of localization in lower dimensions such as the symplectic class, which describes systems with extrinsic spin-orbit interactions\cite{hikamilarkin1980,mirlinreview}. Weak localization (WL), is a powerful complimentary approach, which in contrast to Anderson or strong localization, are its precursor effects arising from phase coherent back-scattering\cite{chakravartyreview} in the metallic regime. 

Following its early development, there has been much progress in understanding how the quantum phase coherence is affected by fundamental processes such as extrinsic and intrinsic spin-orbit interactions, spin-flip scattering, electron-phonon interactions, magnetic fields, Nyquist noise etc....\cite{chakravartyreview} These earlier cases demonstrated a renormalization of the diffusion constant, which can be realized in important phenomena such as anomalous magneto-conductance, anti-localization and spin-selective localization\cite{spinselective,hikamilarkin1980,leerama1985}.
However the existing theoretical studies have not successfully addressed the consequences of non-equilibrium conditions, strong interaction ground states or non-Markovian processes. Therefore it is our purpose here to initiate an exploration of transport paradigms beyond the ordinary case of simple diffusion, by including certain observed features of strongly correlated groundstates such as possible non-equilibrium conditions and non-Markovian affects.

Interactions have been included in disordered electron systems both from a direct perturbative framework (Hartree Fock + impurity vertex corrections)\cite{altshulerlee1980}, a nonlinear sigma model approach\cite{finkJETP1983,castPRB1986,baranovprb2002} and from an alternative scaling analysis\cite{castPRL1987}. More recently, the experimental discovery of a 2D metal-insulator transition (2DMIT) in high mobility SI-MOSFET hetero-structures has received much attention lately\cite{kravRMP01}. Following its discovery, a 2 parameter scaling $\beta$ function which takes the limit of a large valley degeneracy $n_v\to \infty$ in a $1/n_v$ expansion has yielded important and favorable results in understanding transport in dirty Fermi liquid metals\cite{punnoosescience,2007NatPhy}. 

Experimental evidence suggests that the transition point of the 2DMIT takes place within a negative compressibility regime near $r_s \approx 10$ for a wide variety of electron and hole carrier systems\cite{dultzPRL2000}. It should be noted that the neutralizing background renders the total compressibility positive, however this contribution is subtracted off and the density-density correlation function in the limit $q\to0$ for the electron gas itself is negative. In addition, it is well known that strong correlation effects may lead to negative compressibility ground states, although the proper fixed point of the correlated fluid in 2D has not been established. A. Schakel has shown that such conditions may lead to analogues of charge density wave formation and droplet clustering consistent with local density inhomogeneities or imbalance\cite{schakelPRB2001}. Therefore with these experimental and phenomenological conditions in mind, and as an alternative approach we formulate the problem as a reaction-diffusion diffusion process, predicting the destruction of phase coherence and a possible description of non-Fermi liquid groundstates in the negative compressibility regime, not captured by the ordinary Boltzmann picture.

Another interesting case of unconventional transport involves non-Markovian effects,  which is a manifestation of correlated disorder. Earlier studies have detected memory dependent effects by applying an external magnetic field to a 2D electron gas, with observable consequences in the transport\cite{dmitrievPRL2002, dmitrievPRB2008,cheianovPRB2003}. As another application of the semiclassical formalism we include infinite repulsion of the time reversed interference trajectories which can be incorporated into the WL corrections through the scale dependence of the effective scattering length $R(t)$. In simple terms the scattered electrons will remember their past and avoid taking trajectories already traversed, known as the extreme limit of repulsion within the random walk called self avoidance.  

\begin{figure}
\centerline{\includegraphics[width=3.0in]{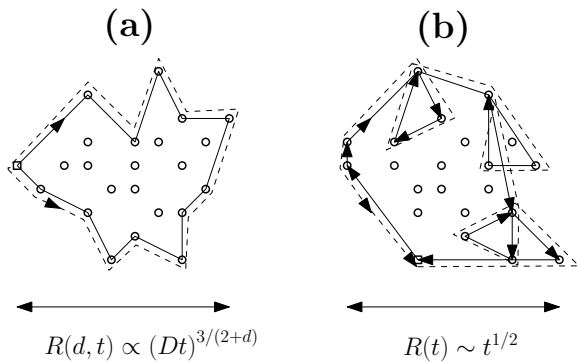}}
\caption{Typical back-scattered trajectories and the expected time dependent behavior of their end to end distances $R(t)$. (a) In a system containing self avoidance none of the paths taken by the electrons may be taken twice, causing the effective scattering length to be dependent on the dimensionality. (b) Ordinary diffusion, where each scattering event is statistically independent of previous ones, allowing for the crossing or retracing of steps.  }
\label{fig:avoid}	
\end{figure}

\emph{Semi-classical weak localization}--
WL can be understood semi-classically in terms of the diffusive behavior of a particle in a $d$-dimensional disordered system. Typically, when the disorder is weak, the mean free path $l$, is much greater than the quantum-mechanical wavelength $\lambda_F \propto k_F ^{-1}$, $k_F l >> 1$ such that the conductivity is modified by,
\begin{equation}
\sigma  = {\sigma _0} + \delta \sigma ,\,\,\,\,\,\,\,\left| {\delta \sigma } \right| \ll {\sigma _0}\,\,\,\,\,\,\,\,\,\,\
\end{equation}
with the usual Drude conductivity given by ${\sigma _0} = {e^2}\rho \tau /{m_e}$ for  charge $e$, carrier density $\rho$, scattering time $\tau$ and mass $m_e$. It is generally appreciated that WL is a quantum-interference contribution to $\delta\sigma$ that results from phase-coherent back-scattering. 
Qualitatively, the quantum interference corrections can be understood from within a Feynman path integral description\cite{chakravartyreview,volldhardtreview}. The Feynman paths (indices $i,j$) connecting two locations within the disordered metal can be described by probability amplitudes $A_i$. The total probability to reach the two points $W$ must sum all possible weighted trajectories, including a cross interference term $W = {\left| {\sum\nolimits_i {{A_i}} } \right|^2} = \sum\nolimits_i {{{\left| {{A_i}} \right|}^2}}  + \sum\nolimits_{i \ne j} {{A_i}A_j^*} $. A back-scattered trajectory that interferes with its time reversed path is quantum mechanically enhanced and facilitates localization\cite{chakravartyreview}.

As a practical matter, in order to calculate $\delta\sigma$ we consider some variant of a diffusive transport along a closed trajectory, fixing the particle's classical return probability $C_D(\vec r,t)$\cite{volldhardtreview}, 
\begin{equation}
\frac{{\partial C_D}}{{\partial t}} = D{\nabla ^2}C_D +\psi(C_D,\vec r,t) 
\label{eq:moddiff}
\end{equation}
where $\psi(C_D,\vec r,t)$ is a function that depends on the specific Hamiltonian considered. Note that $C_D$ is expected to be a normalized probability distribution such that non-singular averages of length $\left\langle x \right\rangle ,\left\langle {{x^2}} \right\rangle $ can be computed. Therefore there should be present a proper competition of terms (growth and dissipation) in $\psi(C_D,\vec r,t)$. Furthermore, in systems with density inhomogeneities and subsequently nontrivial $\psi(C_D,\vec r,t)$, it is not essential for maintaining a consistently normalized Cooperon that the flow of matter in and out of a local space be equal.

Upon determining $C_D(\vec r,t)$ the $\delta \sigma$ correction is extracted from classical considerations of a compact volume $R^d$(over the back-scattered region), spreading as a function of time. $R(t)$ is a central quantity in the study of random walks, which can accurately describe the discrete scattering events of the metallic regime.  In the continuum limit, the random walk becomes a diffusive process controlled by $C_D(\vec r,t)$. Therefore, $R(t)$ is determined from $C_D(\vec r,t)$ by the relation,
\begin{equation}
R(t) = {\left( {\int {{d^d}\vec r{C_D}(\vec r,t){r^2}} } \right)^{1/2}}
\label{eq:rdef}
\end{equation}
which is essentially the root mean squared value of the position taken by the electron. To determine the full contribution to $\delta\sigma$, $R(t)$ must be integrated in the following manner,

\begin{equation}
\delta \sigma  =  - \frac{{  2{e^2}}}{{\pi \hbar }}D\int_\tau ^{\tau _{\varphi}} {dt R(t)^{-d}} 
\label{eq:ds}
\end{equation}
where the probability of return can be expected to be damped $R{(t)^{ - d}} \to R{(t)^{ - d}}{e^{ - (t/{\tau _\varphi })}}$. We therefore include 
$1/\tau_{\varphi}$ as the dephasing rate that regularizes the integration. The interesting case is given by $\tau_{\varphi} \gg \tau$.
Note that the integral in Eq.(\ref{eq:ds}) is precisely the result of the maximally crossed diagrams or Cooperon-like terms generated in the formal Green's function treatment. For the ordinary case of non-interacting electrons in zero external fields, $\tau_{\varphi} \simeq \tau_{ie}$ for a given inelastic scattering time $\tau_{ie}$ and $\psi(C_D,\vec r,t)=0$, yielding
$C_D(\vec r,t) =  e^{- {{\left| {r - {r_0}} \right|}^2}/4Dt} /{\left( {4\pi Dt} \right)}^{d/2}$
leading to $R(t) = \sqrt{2Dt}$ and the conventional WL result,
\begin{equation}
\frac{{\delta \sigma}}{{\sigma_0}}  \propto  \left\{ \begin{array}{l}
 {-\left( {{\tau _{ie}}/\tau } \right)^{1/2}}\,\,\,\,\,\,\,\,\,\,\,\,\,\,d = 1 \\ 
 -\hbar \ln \left( {{\tau _{ie}}/\tau } \right)\,\,\,\,\,\,\,\,\,\,\,d = 2 \\ 
 {-\hbar ^2}{\left( {{\tau _{ie}}/\tau } \right)^{1/2}}\,\,\,\,\,\,\,\,\,d = 3 \\ 
 \end{array} \right.
 \label{eq:wlcorr}
\end{equation}
After restoring the scale dependence of the $\tau$'s by $R(t)$ we have,
\begin{equation}
\sigma (L) = {\sigma _0} - \frac{{{e^2}}}{{\hbar {\pi ^d}}}\left\{ \begin{array}{l}
 (R({\tau _{ie}}) - R(\tau ))\,\,\,\,\,\,\,\,\,d = 1 \\ 
 \ln \left( {R({\tau _{ie}})/R(\tau )} \right)\,\,\,\,\,\,d = 2 \\ 
 \left( {\frac{1}{{R({\tau _{ie}})}} - \frac{1}{{R(\tau )}}} \right)\,\,\,\,\,\,d = 3 \\ 
 \end{array} \right.
\end{equation}

Evidently for $R(\tau _{ie})\gg R(\tau)$, $\sigma (L)$ negatively diverges in $d=1,2$, indicating that the back-scattering contribution no longer has a perturbative effect on the $\sigma$ and such states are strongly localized. If one performs an expansion of the scaling function $\beta(g) = dlng/dlnR(\tau_{ie}) $ near $\epsilon = d-2$ for $\epsilon \ll 1$, what follows is $\beta(g) = d-2-1/g +\dots$ or the celebrated scaling theory of localization\cite{gangfour}.

In general, $\tau_{ie}$ is a functional of R(t) and depends on the power spectrum of the environmental fluctuations. 
Experimentally, the result above can be detected in the limit of low temperatures $T\to0$, given that $\tau_{ie}$ is expected to vanish as a power law in $T$, or $\tau_{ie} \propto T^{p}$. Consequently, in a Fermi-liquid a carefull determination of $p$ is especially required in $d=2$ where the obvious logarithmic correction must be distinguished from the additional logarithmic correction in $T$ generated by the exchange diagrams of the disordered-interacting problem\cite{altshulerlee1980}. This can be accomplished by applying a small normal magnetic field to suppress WL effects\cite{leerama1985}.  Nevertheless, we have provided a self-contained description of the necessary mathematical formalism, indispensable for the rest of the paper.

\emph{Interactions and Reaction-Diffusion}--
As we have shown ordinary diffusion with $R(t) \sim {t^{1/2}}$ produces the WL corrections (\ref{eq:wlcorr}). We strongly emphasize that an alternative way of modeling a different groundstate can be accomplished by including a non-trivial $\psi(C_D,\vec r,t)$ in Eq.(\ref{eq:moddiff}) with possible non-linear and non-equilibrium terms. The generalized Fisher's equation which contains $\psi(C_D,\vec r,t)\propto C_D(1-C_D)$ is the simplest case of reaction-diffusion model, introducing local energy penalties or self interaction for $C_D$ coupled with competing growth terms, often useful in modeling problems related to quantitative biology and physical chemistry in addition to non-equilibrium electronic systems\cite{fisher1937}. In order to get an accurate sense of the scale dependent $\delta\sigma$ and metal-insulator criticality, one needs to determine specific forms of $R(t)$ in arbitrary dimensions, which can be difficult and is itself an active area of research. 

An approximate asymptotic solution of the Fisher's equation in arbitrary dimensions for a wide range of boundary conditions was performed by S. Puri, K.R. Elder and R.C. Desai\cite{asymptfisher}, who determined the following expression for the spreading of the domain,
\begin{equation}
{R_F}(d,t) \sim 2t{\left( {1 - \frac{{d\ln (4\pi t)}}{{2t}}} \right)^{1/2}}
\label{eq:rfisher}
\end{equation}
for a fixed value of $c_F=2$, which is the velocity of the traveling wave solution of $C_D$. $R_F(d,t)$ takes on real values beyond the branch point of the squareroot at ${t_0} =  - (d/2){W_L}\left[ { - 1/(2d\pi )} \right]$, where $W_L$ is the Lambert-W function. At very small values of $t$, $R(t)$ is dominated by the logarithm, however at $t>>t_0$, which is the limit of interest, $R(t)\sim t$, and the $\delta\sigma$ via (\ref{eq:ds}) become,
\begin{equation}
\frac{{\delta {\sigma _F}}}{{{\sigma _0}}} \propto \left\{ \begin{array}{l}
  - \ln (R({\tau _{\varphi}}))\,\,\,\,\,\,\,\,\,\,d = 1 \\ 
  - 2\hbar {(R({\tau _{\varphi}}))^{ - 1}}\,\,\,\,\,d = 2 \\ 
  - 3{\hbar ^2}{(R({\tau _{\varphi}}))^{ - 2}}\,\,\,d = 3 \\ 
 \end{array} \right.
\end{equation}
which is clearly non-singular for large $\tau_{\varphi}$ and contains non-localized states in $d=2$ for the weak disorder limit. This result is rather unexpected, given that we did not introduce any additional scaling couplings, rather a simple unweighted quadratic term in $\psi(C_D,\vec r,t) \sim (C_D)^2$ is sufficient to circumvent localization.

\emph{Interactions, Flory Scaling and Non Markovian Transport}--
Another unconventional groundstate having memory dependent effects with generalized solutions of $R(t)$ in arbitrary dimensions can be constructed by enforcing self avoidance onto the random scattering trajectories. Self avoidance will non-trivially introduce long range correlations into the impurity scattered trajectories of the electrons in contrast to directly considering correlations in the random potential itself. Although there might be a relationship between the two types of correlated randomness, here we are motivated to include repulsion between the trajectories in order to induce the site or path of a scattered electron to have a dependence on its previous scattering history\cite{joebook}.  Physically, a local repulsive term can be introduced into a random walk such that the walk never crosses itself and is free of intersecting loops as shown in Fig.\ref{fig:avoid}(a). Such a model was pioneered by the Nobel prize winning chemist Paul Flory, who developed the famous scaling relation of a random walk with self avoidance\cite{flory1956},
\begin{equation}
{R_{s.a.}}(d,t) \propto {\left( {Dt} \right)^{3/(2 + d)}} = \left\{ \begin{array}{l}
 \left( {Dt} \right)\,\,\,\,\,\,\,\,\,\,d = 1 \\ 
 {\left( {Dt} \right)^{3/4}}\,\,\,\,d = 2 \\ 
 {\left( {Dt} \right)^{3/5}}\,\,\,\,d = 3 \\ 
 \end{array} \right.
\label{eq:fscale}
\end{equation}
where the exponent associated with the spreading of the electron density will have an explicit dependence on the dimensions of the system.
If we introduce Eq.(\ref{eq:fscale}) into (\ref{eq:ds}) we have
\begin{equation}
\frac{{\delta {\sigma _{s.a.}}}}{\sigma } \propto \left\{ \begin{array}{l}
  - \ln \left( {{\tau _{\varphi}}/\tau } \right)\,\,\,\,\,\,\,\,\,\,d = 1 \\ 
  - \hbar {\left( {{\tau _{\varphi}}/\tau } \right)^{ - 1/2}}\,\,\,\,\,\,d = 2 \\ 
  - {\hbar ^2}{\left( {{\tau _{\varphi}}/\tau } \right)^{ - 4/5}}\,\,\,\,d = 3 \\ 
 \end{array} \right.
\end{equation}
\begin{figure}
\centerline{\includegraphics[width=3.0in]{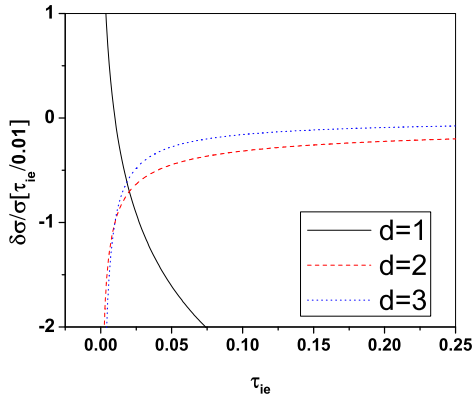}}
\caption{(Color Online) Conductivity corrections in a non-Markovian transport process arising from self-avoidant diffusive scattering. Notice that the logarithmic behavior in $d=1$ is a clear indication of the lower critical dimension of delocalization for which the perturbative $\beta$ function in a $2+\epsilon$ expansion fails to accurately capture the transition.}
\label{fig:splot}	
\end{figure}
which is plotted in Fig.\ref{fig:splot} and after restoring the scattering length dependence for each value of $d$, our final result for the scale dependent $\sigma$ is given by
\begin{equation}
\begin{array}{l}
 \frac{\delta {\sigma _{s.a.}}}{\sigma } \propto  \left\{ \begin{array}{l}
 \ln (R(1,{\tau _{\varphi}})/R(1,\tau ))\,\,\,\,\,\,\,\,\,\,\,\,\,\,\,\,\,\,\,\,\,\,\,\,\,\,\,\,\,\,\,\,\,\,d = 1 \\ 
 2\left( {{R^{ - 2/3}}(2,{\tau _{\varphi}})\, - \,{R^{ - 2/3}}(2,\tau )} \right)\,\,\,\,d = 2 \\ 
 \frac{5}{4}\left( {{R^{ - 4/3}}(3,{\tau _{\varphi}})\, - \,{R^{ - 4/3}}(3,\tau )} \right)\,\,\,\,d = 3 \\ 
 \end{array} \right. \\ 
 \end{array}
\end{equation}
showing exclusive localization only in $d=1$, and the possibility for a mobility edge in 2D now emerges near the band center. Although a 2DMIT is expected at some critical value of the ratio of random potential width over the hopping strength, it is necessary for accurate numerical simulations using the finite size scaling method to confirm these aspects of the transition, and determine the critical exponents.

\emph{Discussion and concluding remarks}--
For both cases studied here, we have observed a failure of localization in $d=2$ simply by restricting the back-scattering by means of a modified diffusion process and determining its perturbative effect on the transport. These two new types interactions in disordered electron systems may reflect an alternative ground state to the conventional Fermi liquid. First we have calculated the transport corrections in a Fisher's reaction- diffusion model, where the interactions are present as nonlinearities in the diffusion (or self interaction of the probability density) rather than directly taken as a density-density (or four fermion) interaction in the original Hamiltonian. 

Thus we have chosen as an alternative to a many-body picture, a one particle system with modified kinetics, which reflects a possibly different fixed point of the electron liquid. Furthermore, in the Fisher's equation there is a local repulsion of the probability density in addition to non-equilibrium growth which may be an emergent manifestation of the strongly correlated electron liquid, as both theoretical and experimental evidence suggests that there may be density inhomogeneities, negative compressibility states\cite{dultzPRL2000,schakelPRB2001} near the 2DMIT and the predicted nematic phases that are intermediate between the Fermi-liquid and the Wigner crystal\cite{kivelsonspivak}. Hence, experimentally, by tuning the density of a 2DEG in a disordered SI-MOSFET heterostructure in the vicinity of $r_s\approx10$ near the 2DMIT critical point we expect the negative compressibility conditions modeled by the phenomenological considerations of a reaction-diffusion transport model to hold accurately. 

Regarding non-Markovian transport, recent $\sigma(T)$ measurements in disorder tuned magnetic films have detected regimes with pure power-law like corrections to $\sigma(T)$\cite{ghosh2010APS}. Additionally ferromagnetic semiconducting systems are known to contain non-Markovian transport and spin relaxation mechanisms for which anomalous temperature dependent corrections have been experimentally observed in $d=1,2$\cite{glazovEPL2006,dietlJPSJ2008}. Hence, materials such as ferromagnetic (Ga,Mn)As are likely candidates for observing the effects presented here.

In conclusion, we have identified several unconventional modes of diffusive transport that can circumvent localization and it is shown here for the first time that non-equilibrium conditions can destroy the localization corrections in 2D. We attempted to develop a discussion of the single parameter scaling of $\beta(g)$ which should depend on the special symmetries of the Hamiltonian. Hence, as a compliment to our semiclassical scaling theory it is essential that several possible microscopic Hamiltonians and stable fixed points that reflect the environmental conditions for these forms of restricted backscattering can be properly reconstructed and those relevant symmetries are identified in addition to careful numerical simulations which are beyond the valid perturbative regime of our analysis.

\begin{acknowledgments}
I am grateful for general discussions about localization with Y. Tserkovnyak, K. Slevin and A. Furusaki and I thank the Akai Laboratory at Osaka university for its hospitality and where some of this work was conceived.
This work was supported by the Institute for Physical and Chemical Research (RIKEN) FPR postdoctoral fellowship. 
\end{acknowledgments}

\end{document}